# Temperature dependence of Magnetic properties in Nanocrystalline copper ferrite thin films

Prasanna D. Kulkarni[1], Shiva Prasad[1], N. Venkataramani[2], R. Krishnan[3], Wenjie Pang[4], Ayon Guha[4], R. C. Woodward[4], R.L. Stamps[4]

[1]Physics Department, IIT Bombay, [2]ACRE, IIT Bombay, Mumbai, 400076, [3]Laboratoire de Magnetism et d'optique de Versailles, CNRS, 78935, France, [4]School of Physics, M013, The University of Western Australia, 35 Stirling Hwy, Crawley WA.

*Abstract*

*The copper ferrite thin films have been deposited by RF sputtering at a 50W rf power. The As-deposited films are annealed in air at 800ºC and then slow cooled. The As-deposited (AD) as well as slow cooled (SC) films are studied using a SQUID Magnetometer. The M Vs H curves have been recorded at various temperatures between 5K to 300K. The coercivities obtained from the MH curves are then plotted against temperature (T). The magnetization in the films does not saturate, even at the highest field of 7T. The high field part of the M Vs H curves is fitted using the $H^{1/2}$ term of Chikazumi expression $M(H)= Q*(1- a/H^n)$, with n=1/2. The variation of coefficient 'a' of $H^{1/2}$ term has been observed with temperature (T). An attempt has been made to correlate this with the coercivity (Hc) in the case of annealed films.*

## INTRODUCTION

The copper ferrite can be stabilized in two different phases in thin film form, viz., a cubic and a tetragonal phase. The cubic phase is stabilised in the as deposited (AD) films. The slow cooling (SC) of the films after ex-situ annealing results in the tetragonal phase[1]. It has also been observed that the RF sputtered films are nanocrystalline in nature[1]. The magnetic measurements of the nanocrystalline film shows non saturation of the MH loop even at high fields. This is termed as High Field Suceptibility (HFS). In the case of rf sputtered nanocrystalline ferrite thin films, the expression, $M(H)= Q*(1- a/H^n)$ fitted the approach to saturation best with n=1/2 [2]. The coefficient 'a' of the $H^{1/2}$ term is attributed to point like defects present in the material. In the present paper, the temperature (T) dependence of coercivity ($H_c$) has been discussed for both cubic and tetragonal phases of the Cu ferrite films. The variation of $H_c$ with T is also compared with the variation of coefficient 'a' of $H^{1/2}$ term of Chikazumi expression [3].

## EXPERIMENTAL

The copper ferrite films were deposited using the Leybold Z400 rf sputtering system. The films were deposited on amorphous quartz substrates at ambient temperature during sputtering. The rf power employed during the deposition is 50W. The thickness of the films are ~2400 Å. The AD films were ex-situ annealed in air at 800°C for 2 hours, followed by slow cooling. The MH loops were measured for AD and SC films in a field up to 7T using a SQUID magnetometer at various temperatures between 5K and 300K.

## RESULTS

Fig. 1 shows the Hc Vs T data for AD and SC samples of 50W copper ferrite films. At all temperatures, the $H_c$ values for AD film are lower than the SC film. For the AD film, $H_c$ varies from 35 Oe at 300K to 2210 Oe at 5K. For the SC film also coercivity increases with decreasing temperature. However, the variation is over a smaller range from 800 Oe at 300K to 2300 Oe at 5K. Moreover the functional form of the variation of the coercivity appears to be quite different for the two cases.

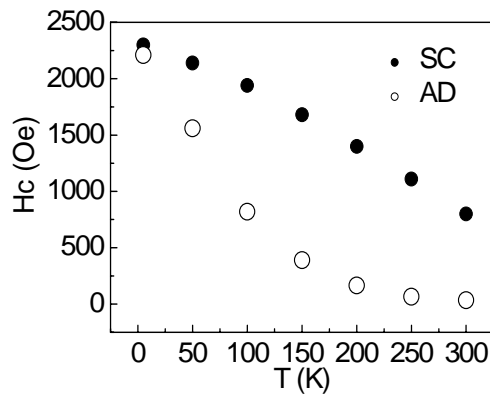

**Fig 1 :** Temperature dependence of Hc for 50W, AD and SC Cu ferrite films.

## DISCUSSIONS

The approch to saturation was discussed by Chikazumi [3]. In the Chikazumi expression, the mangetization at a given applied field is written in the following way,

$$4\pi M = Q * (1 - a/H^{1/2} - b/H - c/H^2 - ....) + eH \qquad (1)$$

Here $4\pi M$ is the actual value of magnetization that is observed at a field H, and Q, a, b, c, and e are constants. The value of Q should correspond to a value of $4\pi M$ in infinite field. The second term, $a/H^{1/2}$ is attributed to the point like defects or magnetic anisotropy fluctuations on atomic scale. We have fitted the magnetization data from 0.8T field to 7T field to an expression involving just $H^{1/2}$ term in eq. (1). The value of '$a$' have been obtained at various temperatures between 5K and 300K.

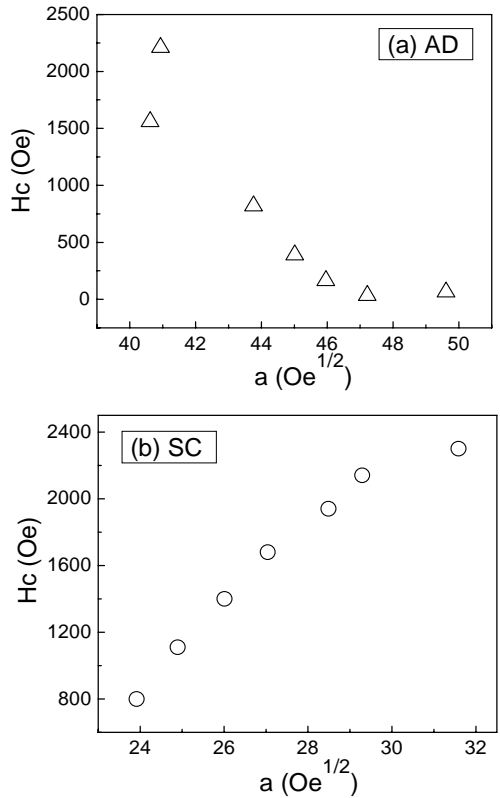

**Fig 2 :** The variation of $H_c$ Vs '$a$' for 50W, (a) AD and (b) SC, Cu ferrite films.

The $H_c$ and '$a$' values at various temperatures are plotted against each other in Fig. 2. The $H_c$ values are plotted in the increasing order along y-azis, which is equivalent to decreasing T. For the SC film, the $H_c$ and '$a$' values both increase with decrease in T. While for AD film, the '$a$' value decreases and $H_c$ increases with decreasing T. For AD film, the variations of $H_c$ and '$a$' with temperature are opposite. If anisotropy is the only parameter controlling both $H_c$ and '$a$', their values are expected to increase with decrease in temperature. This is not observed for AD film.

The grain sizes in the case of rf sputtered AD films, have been observed to be ~5 – 10 nm [1]. In this range, the grains may become superparamagnetic [4]. However, if superparamagnetism is dominant in AD films at lower temperatures, the value of '$a$' is expected to increase, which is not observed. It is also possible that decrease in '$a$' with lowering T in AD films is because of blocking of superparamagnetic particles at lower temperatures.

## CONCLUSIONS

The variation of $H_c$ and '$a$' with temperature shows opposite behaviour for AD rf sputtered films. This can not be completely understood on the basis of anisotropic effects or superparamagnetism.

## ACKNOWLEDGEMENT

The author Prasanna D. Kulkarni acknowledge the CSIR, India for financial support.